\documentclass[a4paper,11pt]{article}
\usepackage{pos}
\usepackage{media9}

\DeclareMathOperator{\Tr}{Tr}

\title{SU(3) centre vortex geometry at finite temperature}

\author{Jackson A. Mickley}
\author{Waseem Kamleh}
\author*{Derek B. Leinweber}
\affiliation{Centre for the Subatomic Structure of Matter, Department of Physics, The University of Adelaide, South Australia 5005, Australia}

\emailAdd{jackson.mickley@adelaide.edu.au}
\emailAdd{derek.leinweber@adelaide.edu.au}

\abstract{The importance of examining the structure of centre-vortex matter in the ground-state
  fields of nonabelian gauge-field theory has been demonstrated in the recent centre-vortex based
  discovery of a second finite-temperature transition in QCD associated with quark
  deconfinement. This signals the presence of a new phase of ground-state field structure between
  the well separated chiral and deconfinement transitions.
  In this short presentation, we re-examine pure SU(3) gauge theory which provides a foundation for
  the development of techniques for the examination of full QCD.  This time, we reconsider
  visualisations of the centre-vortex structure in light of the quantitative analysis that
  demonstrates the first order nature of the deconfinement phase transition in the pure-gauge
  theory.  Here we consider a detailed side-by-side comparison of the field structure slightly
  below and slightly above the critical temperature.  The abrupt changes of the field structure in
  the first order phase transition are easy to observe in the representative visualisations.}

\FullConference{The XVIth Quark Confinement and the Hadron Spectrum Conference (QCHSC24)\\
 19-24 August, 2024\\
 Cairns Convention Centre, Cairns, Queensland, Australia\\}

\begin{document}
\maketitle

\section{Introduction}

Through an examination of centre-vortex matter in the ground-state fields of QCD and the dependence
of its structure on temperature \cite{Mickley:2024zyg}, a second finite-temperature transition
associated with quark deconfinement has been discovered \cite{Mickley:2024vkm}. The discovery of a
new phase of QCD ground-state field structure between the well separated chiral and deconfinement
transitions emphasises the importance of examining centre-vortex structure 
\cite{%
tHooft:1977nqb,tHooft:1979rtg,Nielsen:1979xu,%
DelDebbio:1996lih,Faber:1997rp,DelDebbio:1998luz,Montero:1999by,Bertle:1999tw,%
Faber:1999gu,Engelhardt:1999xw,Bertle:2000qv,Engelhardt:2003wm,%
Langfeld:2001cz,%
Greensite:2003bk,%
Langfeld:2003ev,%
OCais:2008kqh,%
Bowman:2010zr,%
OMalley:2011aa,%
Trewartha:2015nna,%
Trewartha:2015ida,%
Greensite:2016pfc,%
Trewartha:2017ive,%
Biddle:2018dtc,%
Biddle:2019gke,%
Biddle:2020isk,%
Biddle:2022acd,%
Biddle:2022zgw,%
Virgili:2022ybm,%
Biddle:2023lod,%
Kamleh:2023gho,%
Mickley:2025ksp%
} in QCD.  For more recent overviews see Refs.~\cite{Leinweber:2022ukj,Gross:2022hyw}.

The visualisation and animation of centre-vortex structure is helpful in understanding the nature
of vortex matter and vital to the formulation of new measures to quantify its structure. Good
examples of this include the novel vortex persistence correlation measure \cite{Mickley:2024zyg},
the role of secondary clusters as a signature of deconfinement \cite{Mickley:2024zyg}, and the
rigorous definition of a vortex branching probability \cite{Biddle:2023lod}.

In this short presentation, we re-examine pure SU(3) gauge theory \cite{Mickley:2024zyg} which
provides a foundation for the development of techniques for the examination of full QCD
\cite{Mickley:2024vkm}.  This time, we reconsider visualisations of the centre-vortex structure in
light of the quantitative analysis of Ref.~\cite{Mickley:2024zyg} that demonstrates the first order
nature of the deconfinement phase transition in the pure-gauge theory.

Here we consider a detailed side-by-side comparison of the vortex-matter field structure slightly
below and slightly above the critical temperature.  Drawing on the quantitative analysis presented
in the figures of Ref.~\cite{Mickley:2024zyg}, we'll search for signatures of the established
properties in the representative illustrations of the field structure. The abrupt step changes of the
field structure in the first order phase transition are easy to observe in the visualisations and
animations.

We begin with a brief review of centre vortex identification and visualisation techniques in the
following section.  The detailed examination of the field structure follows in
Sec.~\ref{sec:changes} where interactive 3D visualisations and animations are presented.  We close
with a few concluding remarks.

\section{Centre-vortex visualisation techniques}
\label{sec:id}

Physical vortices in the QCD ground-state fields have a finite thickness and so permeate all four
spacetime dimensions. In contrast, on the lattice ``thin'' centre vortices are extracted through a
well-known gauge-fixing procedure that seeks to bring each link variable $U_\mu(x)$ as close as
possible to an element of $\mathbb{Z}_3$.  This is known as Maximal Centre Gauge (MCG). 

Fixing to MCG is typically performed by finding the gauge transformation $\Omega(x)$ to maximise
the functional \cite{Montero:1999by}
\begin{equation}
    R = \frac{1}{V\,N_\mathrm{dim}\,N_c^2} \,\sum_{x,\,\mu} \,\left| \Tr U_\mu^{\Omega}(x) \right|^2 \,, \qquad V = N_s^3 \times N_\tau \,.
\end{equation}
The links are subsequently projected onto the centre,
\begin{equation}
	U_\mu(x) \longrightarrow Z_\mu(x) = \exp\left(\frac{2\pi i}{3} \,n_\mu(x) \!\right) \mathbb{I} \in \mathbb{Z}_3 \,,
\end{equation}
with $n_\mu(x) \in \{-1,0,1\}$ identified as the centre phase nearest to $\arg \Tr U_\mu(x)$ for
each link. Finally, the locations of vortices are identified by nontrivial plaquettes in the
centre-projected field
\begin{equation}
	P_{\mu\nu}(x) = \prod_\square Z_\mu(x) = \exp\left(\frac{2\pi i}{3} \, m_{\mu\nu}(x) \!\right)\mathbb{I}
\end{equation}
with $m_{\mu\nu}(x)$ having nontrivial values $\pm 1$. The value of $m_{\mu\nu}(x)$ is referred to
as the \textit{centre charge} of the vortex, and we say the plaquette is pierced by a vortex.
The centre charge is conserved such that the vortex topology manifests as closed sheets in four
dimensions, or as closed lines in three-dimensional slices of the lattice.

\begin{figure}[tb]
    \centering
    \includegraphics[width=0.6\linewidth]{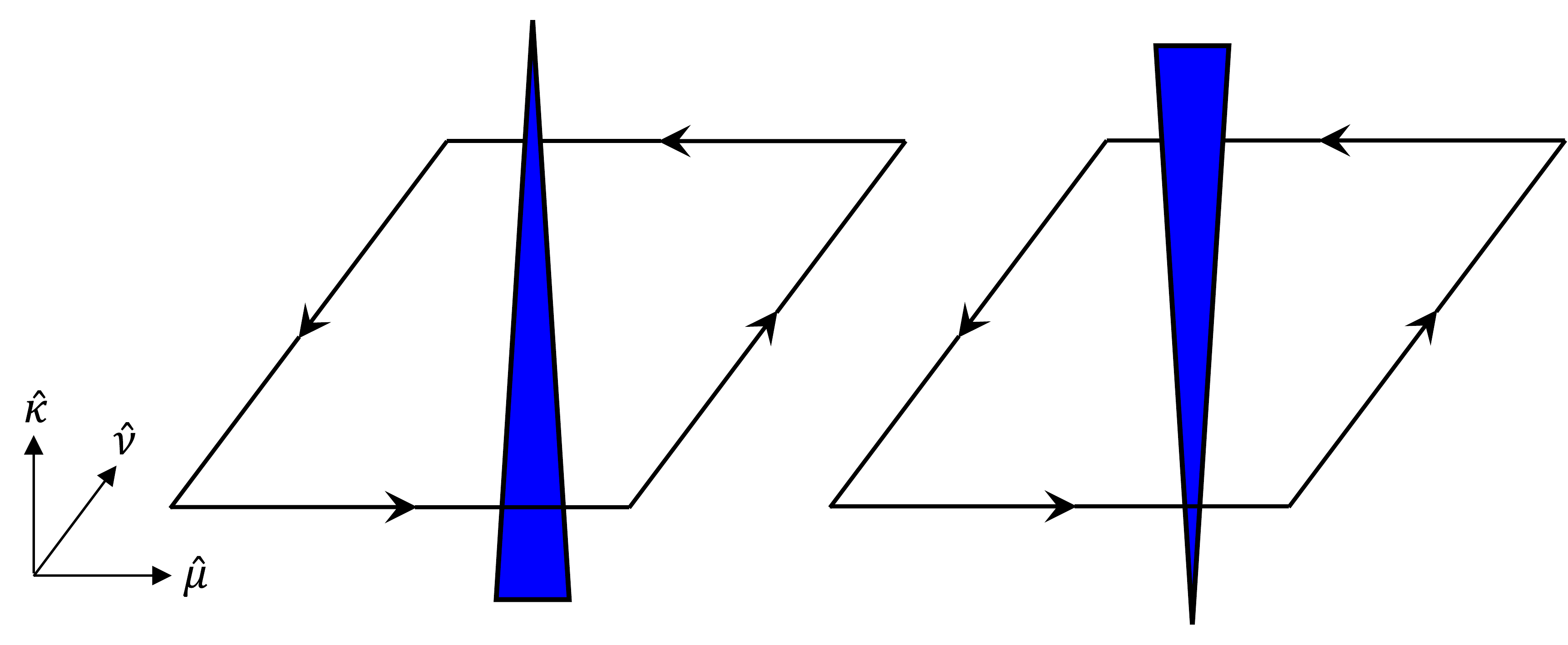}
    \caption{\label{fig:visconvention} The visualisation convention for centre vortices. An $m=+1$
      vortex (\textbf{left}) is represented by a jet in the available orthogonal dimension, with
      the direction given by the right-hand rule. An $m=-1$ vortex (\textbf{right}) is rendered by
      a jet in the opposite direction.}
\end{figure}

With the locations of vortices identified, we draw on the visualisation techniques previously
established in Ref.~\cite{Biddle:2019gke}.
To construct a 3D visualisation we slice on a given dimension.  For example temporal slices have
the time coordinate, $\tau$, fixed, whereas spatial slices have one of the spatial coordinates,
$x$, $y$, or $z$, fixed.
This leaves one direction orthogonal to a non-trivial plaquette that can be used to identify the
sign of the centre charge.  
As vortices exist on the dual lattice, the point-like centre vortex for
\begin{equation} \label{eq:projectedplaquete}
    P_{\mu\nu}(x) = \exp\left(\frac{2\pi i}{3} \,m_{\mu\nu}(x) \right) \,,
\end{equation}
is plotted at the centre of the plaquette.  As the centre charge can take values of $\pm1$ we
render this information as a jet pointing in the third dimension orthogonal to the $\mu$-$\nu$
plane with values of $+1$ ($-1$) pointing in the positive (negative) direction.
As such, the visualisations exclusively show the flow of $m=+1$ centre charge. This convention is
demonstrated in Fig.~\ref{fig:visconvention}.
Finally, utilising the SU(3) cluster identification algorithm developed in
Ref.~\cite{Biddle:2023lod}, the percolating cluster is rendered with blue jets and secondary vortex
cluster loops are rendered in contrasting colours to aid in their identification.

The connection between centre vortices and confinement is apparent through space-time Wilson loops
of size $R\times T$, which asymptotically give access to the static quark potential $V(r)$
\begin{align} \label{eq:staticquarkpotential}
    \langle W(R,T) \rangle \sim \exp\left(-V(r) \,a\,T\right) \,, && \! r=Ra\,, && \! T \text{ large} \,.
\end{align}
The percolation of centre vortices through spacetime at low temperatures provides an area law for
the Wilson loop corresponding to a potential $V(r)$ that linearly rises with the separation $r$.
Thus in examining the centre-vortex structure that gives rise to confinement, it will be essential
to render spatial slices of the lattice such that vortices piercing space-time plaquettes can be
observed.

\section{Changes in vortex-matter structure through the deconfinement phase transition}
\label{sec:changes}

\subsection{Temporal Slices}

\subsubsection{Overview}

We begin by studying the more intuitive temporal slices of the finite temperature lattices of
Ref.~\cite{Mickley:2024zyg} where the temporal coordinate is held fixed and the spatial volume is
visualised.  
The lattices have a fixed isotropic lattice spacing $a=0.1\,$fm and a spatial volume of $32^3$.
The ensembles were generated with the Iwasaki renormalisation-group improved action
\cite{Iwasaki:1996sn}. 

Here we examine two temperatures very close to the critical temperature of the first order phase
transition.  A temporal extent of $N_\tau = 8$ provides $T/T_c = 0.91$ just below the critical
temperature, $T_c$, and $N_\tau = 6$ provides $T/T_c = 1.22$ just above $T_c$.

\subsubsection{Vortex density}

\begin{figure}[p]
\centering
  \includemedia[
  noplaybutton,
  3Dtoolbar,
  3Dmenu,
  label=Nt8temporal,
  3Dviews=graphics/U3D/Nt8_temporal_slice.vws,
  width=12.5cm,
  ]{\includegraphics{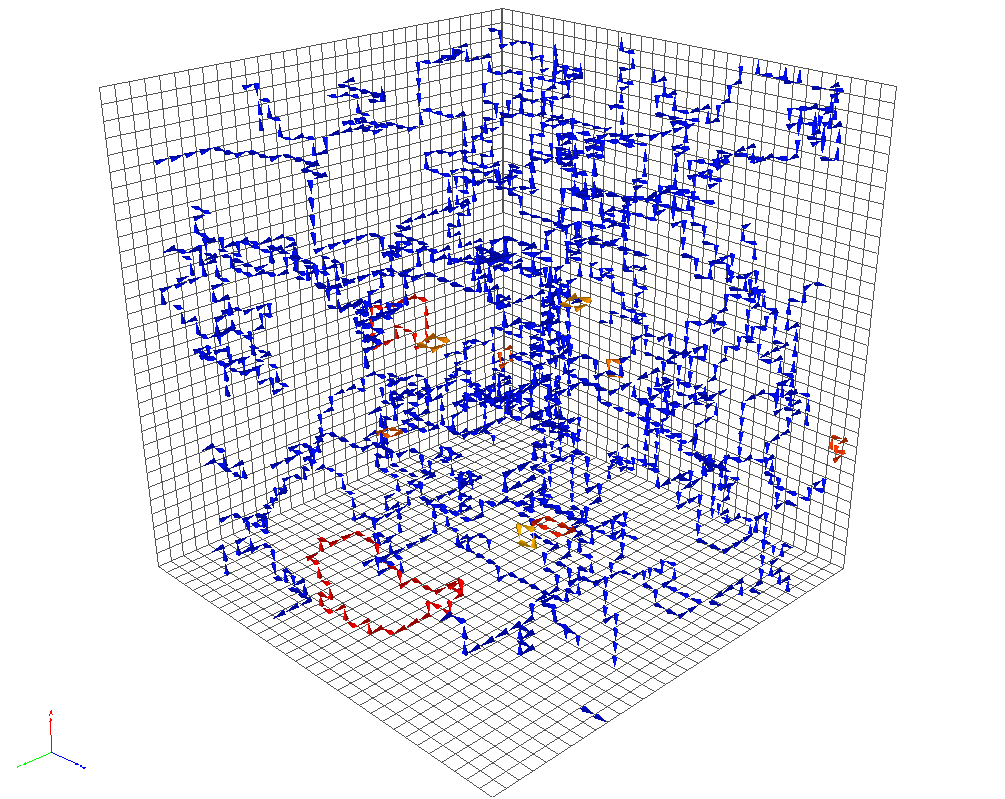}}{graphics/U3D/Nt8_temporal_slice.u3d}

\null\vspace{1pt}

  \includemedia[
  noplaybutton,
  3Dtoolbar,
  3Dmenu,
  label=Nt6temporal,
  3Dviews=graphics/U3D/Nt6_temporal_slice.vws,
  width=12.5cm,
  ]{\includegraphics{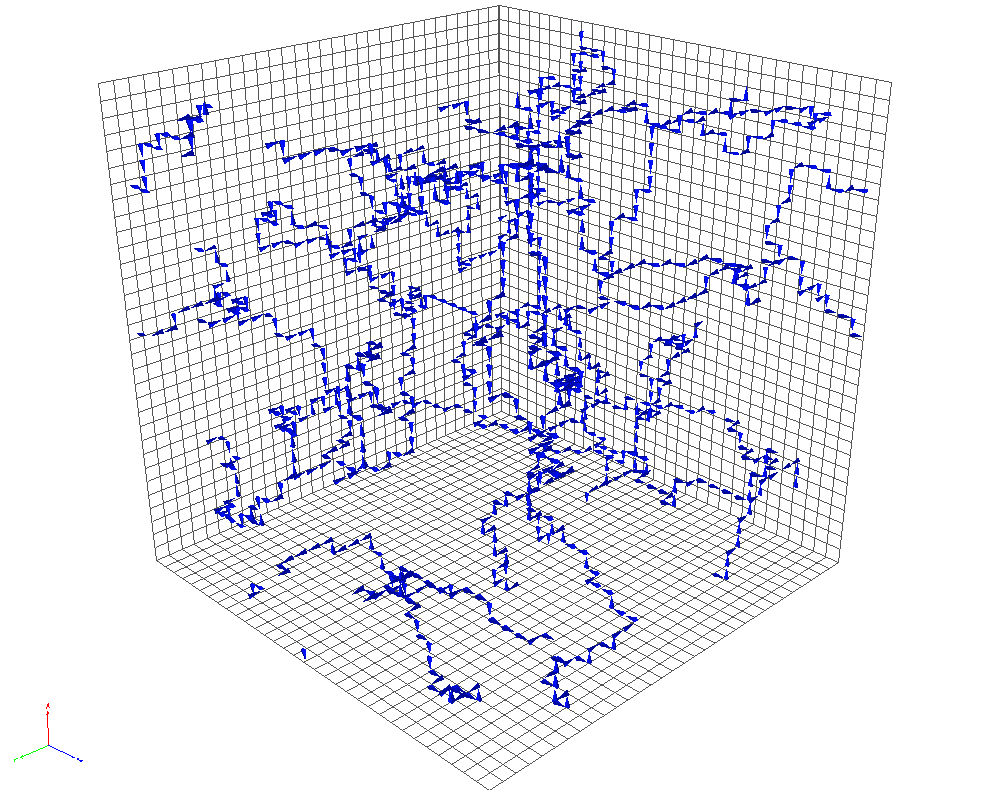}}{graphics/U3D/Nt6_temporal_slice.u3d}
\caption{Illustrations of a temporal slice of the finite temperature lattices with $N_\tau = 8$
  providing $T/T_c = 0.91$ (\textbf{upper}) and $N_\tau = 6$ providing $T/T_c = 1.22$
  (\textbf{lower}).  The blue jets illustrate the nontrivial plaquettes pierced by a vortex forming
  the percolating cluster. Secondary vortex clusters separate from the percolating cluster on this
  time slice are rendered in different colours to aid in their identification. }
\label{fig:tempSlice}
\end{figure}

Figure~\ref{fig:tempSlice} provides an interactive rendering of the 3D structure of the centre
vortex matter.  To activate these illustrations and the animations to come, readers will need to
open this document with Acrobat Reader or Foxit Reader.
Click to activate, click and drag to rotate the image, control-click to translate and shift-click
or mouse wheel to zoom into the image. The Views menu can be used to view specific features
discussed in the text.

Even without activating the figures for closer inspection, one can see significant difference in
the density of vortices.  Above $T_c$ the density drops from $1.8$ to $1.2\,$fm$^{-2}$,
approximately $2/3$ of the original density.  
Figure~10 of Ref.~\cite{Mickley:2024zyg} 
illustrates the first order nature of the phase transition and the evolution of the vortex density
with temperature.

\subsubsection{Secondary clusters}

The next clear signature of a phase transition is the suppression of secondary clusters in 3D
temporal slices of the lattice.  This is a robust deconfinement signature associated with the
alignment of the vortex surface in four dimensions with the temporal axis.

In a study of $SU(4)$ gauge theory \cite{Mickley:2025ksp}, animations of the vortex structure over
temporal slices revealed how these secondary clusters migrate in the fourth dimension, thus
providing a rendering of the two-dimensional vortex sheet in four dimensions.  There one could see
how the secondary clusters could migrate and eventually join the percolating structure.

\begin{figure}[tb]
\centering
\includegraphics[width=0.6\linewidth]{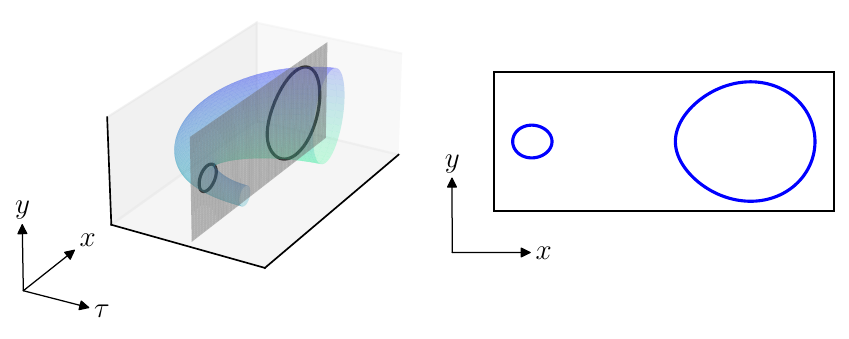}

\vspace{-0.5em}

\includegraphics[width=0.6\linewidth]{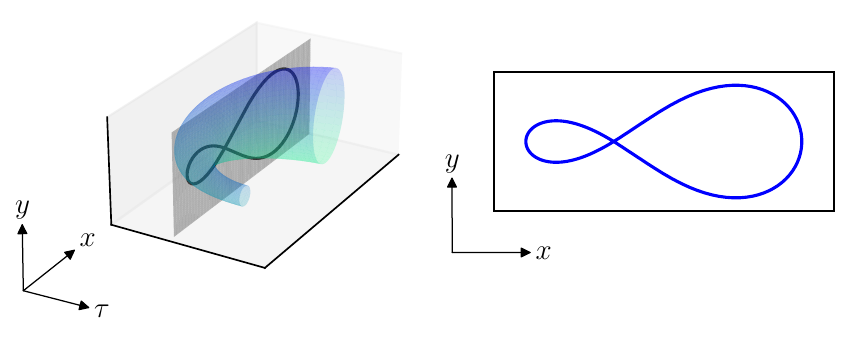}

\vspace{-0.5em}

\includegraphics[width=0.6\linewidth]{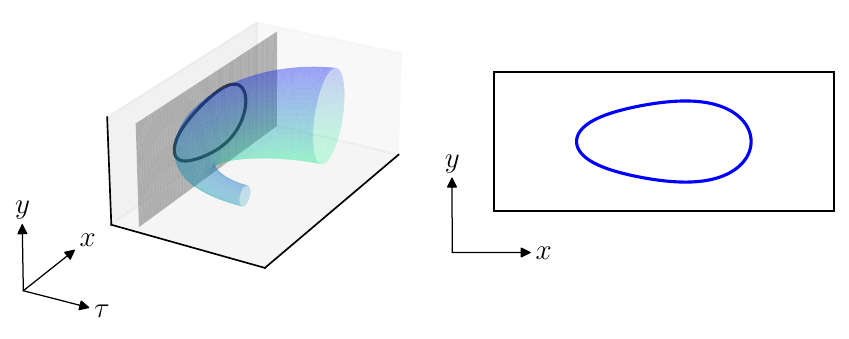}
\caption{\label{fig:secondary} An simplified illustration showing how an apparently disconnected
  secondary cluster in a temporal slice of the lattice can be part of a single vortex sheet in four
  dimensions.
  Here, we are taking the equivalent of a temporal slice of the vortex surface by fixing the time
  coordinate (left column) and looking at an $x$-$y$ cross section (right column).
  In the \textbf{top} illustration, the small loop in the fixed-$\tau$ cross section appears to
  disconnected from the larger loop.  Because the surface curves back on itself in the temporal
  dimension, one can slice the surface at the time where the two loops connect (\textbf{middle}) to
  eventually form a single cluster (\textbf{bottom}). All loops belong to the same vortex sheet.  }
\end{figure}

A simplified illustration of this is provided in Fig.~\ref{fig:secondary}.  The key requirement is
that the vortex sheet curves back on itself in the temporal direction.  The alignment of the vortex
sheet with the temporal direction suppresses this curvature such that the number of secondary
clusters drops significantly across the critical temperature.  In essence, the absence of secondary
clusters is a novel measure of vortex sheet alignment with the temporal direction.
Figure~12 of Ref.~\cite{Mickley:2024zyg} illustrates this significant drop across the phase
transition, and the continued suppression of secondary clusters as one moves to higher densities.

\subsubsection{Branching Points}

\begin{figure}[tb]
\centering
\includegraphics[width=0.3\linewidth]{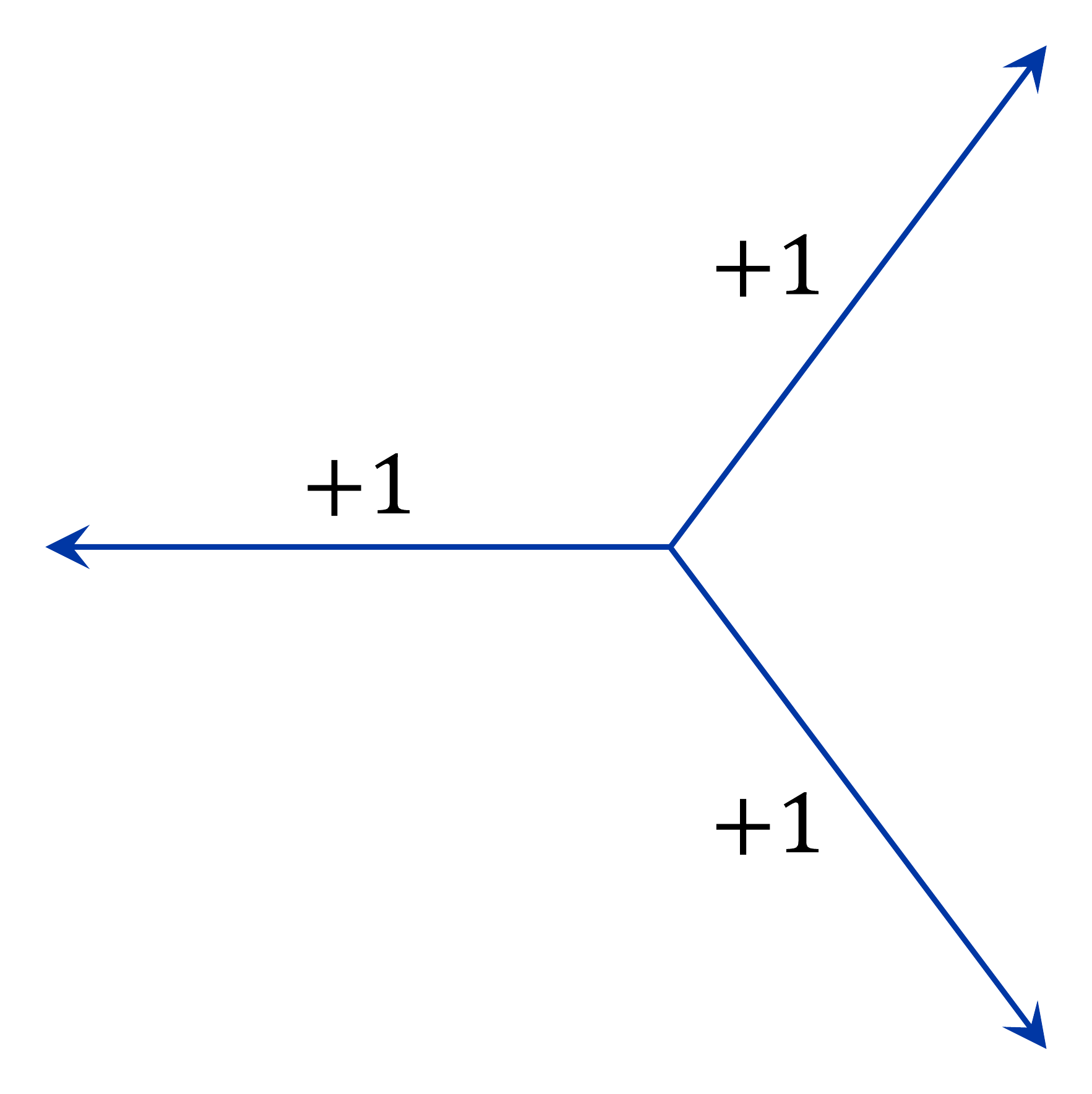}\qquad
\includegraphics[width=0.3\linewidth]{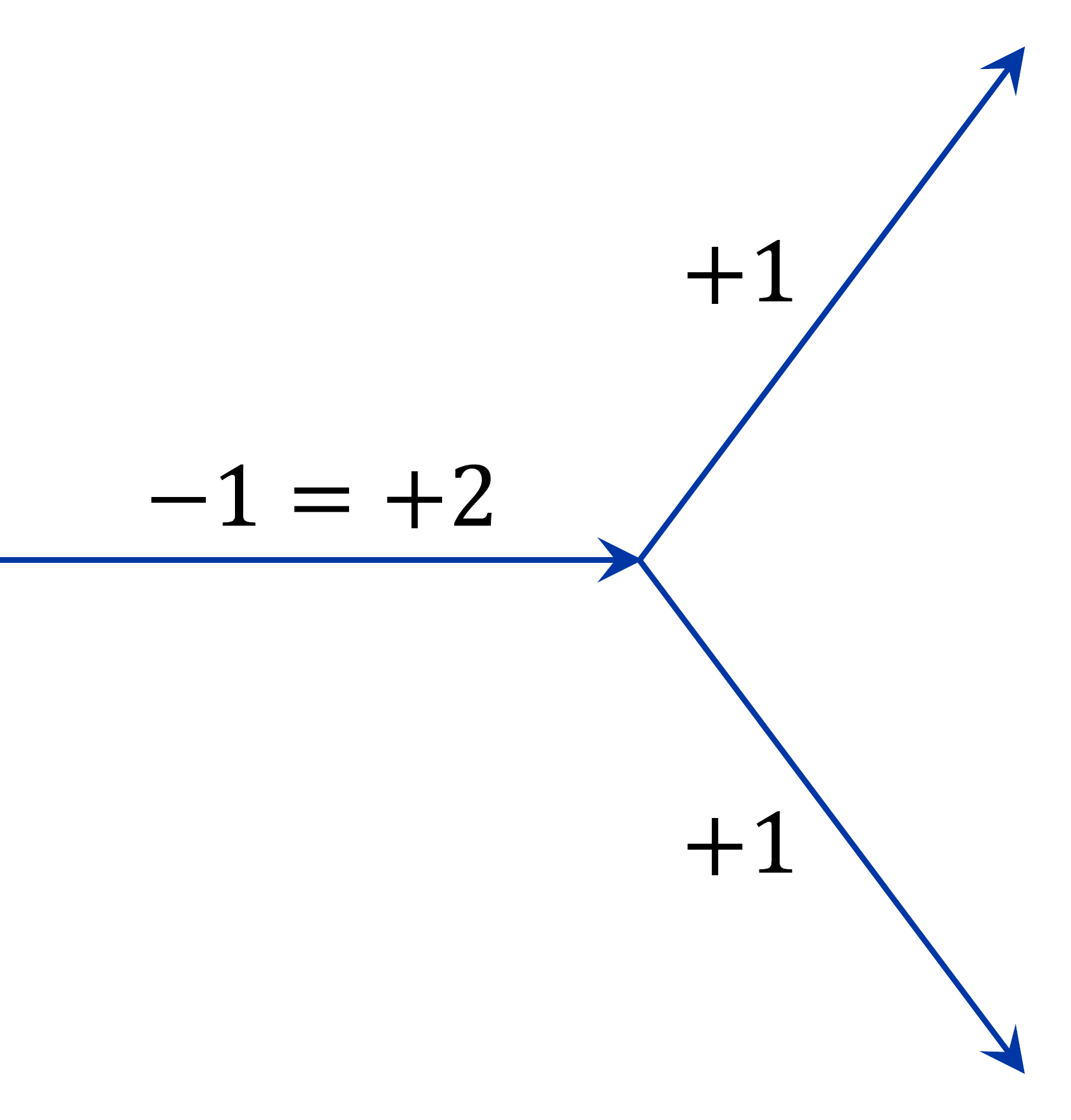}
\caption{Schematic of a monopole vertex (\textbf{left}) versus a branching point
  (\textbf{right}). The monopole vertex follows our convention to illustrate the directed flow of
  $m=+1$ centre charge. Reversal of the left-hand arrow in the diagram shows the flow of charge
  $m=-1$, as seen on the right. Due to periodicity in the centre charge, $m=-1$ is equivalent to
  $m=+2$. Thus, the right-hand diagram depicts branching of centre charge.}
\label{fig:branching} 
\end{figure}

Figure~15 of Ref.~\cite{Mickley:2024zyg} shows a significant drop in the branching point density
across the phase transition and this is also manifest in Fig.~\ref{fig:tempSlice}.  In our
visualisations illustrating the flow of centre charge $+1$, these points appear as (anti-)monopoles
with three jets converging or emerging from a cube of the lattice. Figure \ref{fig:branching}
reminds the reader of the relationship between monopoles and branching points.

While part of the reduction in branching point density can be attributed to a general loss of
vortex density, the lower density is also manifest in the absence of vortex chains above $T_c$
resembling knots.  This feature is related to another characteristic of the first order phase
transition, where there is a change in the tendency of branching points to cluster at low
temperatures.

The suppression of branching-point clustering above $T_c$ is illustrated in the histograms of
vortex chain lengths between branching points in Fig.~17 of Ref.~\cite{Mickley:2024zyg}.  For
temperatures below $T_c$ there is a significant excess in small vortex chain lengths, well above the
exponential fall off for larger chain lengths.  This excess disappears across the phase transition
as most clearly illustrated in the logarithmic vortex chain histograms in Fig.~19 of
Ref.~\cite{Mickley:2024zyg}.

Returning to our temporal slice rendering in Fig.~\ref{fig:tempSlice}, an example of vortex
clustering below $T_c$ is provided in the ``Branching point clustering'' view in the upper
rendering.  There six branching points can been seen within the zoomed-in view. Three of these
branching points are immediately adjacent to each other.  Similar features do not appear above $T_c$.
The view entitled ``Linear vortex chains'' shows an example of how the vortex chains tend to move
away from a branching point in a linear manner without further branching in the immediate
proximity.  

Finally, it is worth noting that the branching point probability also drops across the phase
transition as illustrated in Fig.~20 of Ref.~\cite{Mickley:2024zyg}.  This is a more subtle change
that could not be confidently observed in the representative slices without the quantitative
analysis of Ref.~\cite{Mickley:2024zyg}.

\subsection{Spatial Slices}

\subsubsection{Overview}

\begin{figure}[p]
\centering
  \includemedia[
  noplaybutton,
  3Dtoolbar,
  3Dmenu,
  label=Nt8spatial,
  3Dviews=graphics/U3D/Nt8_spatial_slice.vws,
  width=12.5cm,
  ]{\includegraphics{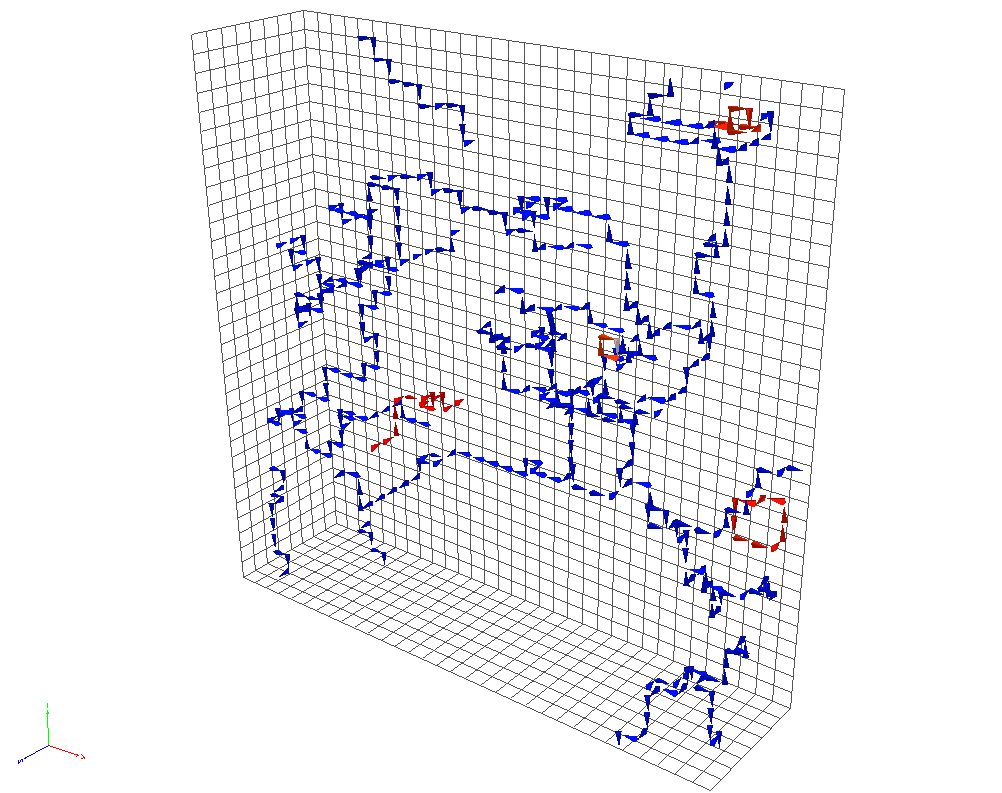}}{graphics/U3D/Nt8_spatial_slice.u3d}

\null\vspace{1pt}

  \includemedia[
  noplaybutton,
  3Dtoolbar,
  3Dmenu,
  label=Nt6spatial,
  3Dviews=graphics/U3D/Nt6_spatial_slice.vws,
  width=12.5cm,
  ]{\includegraphics{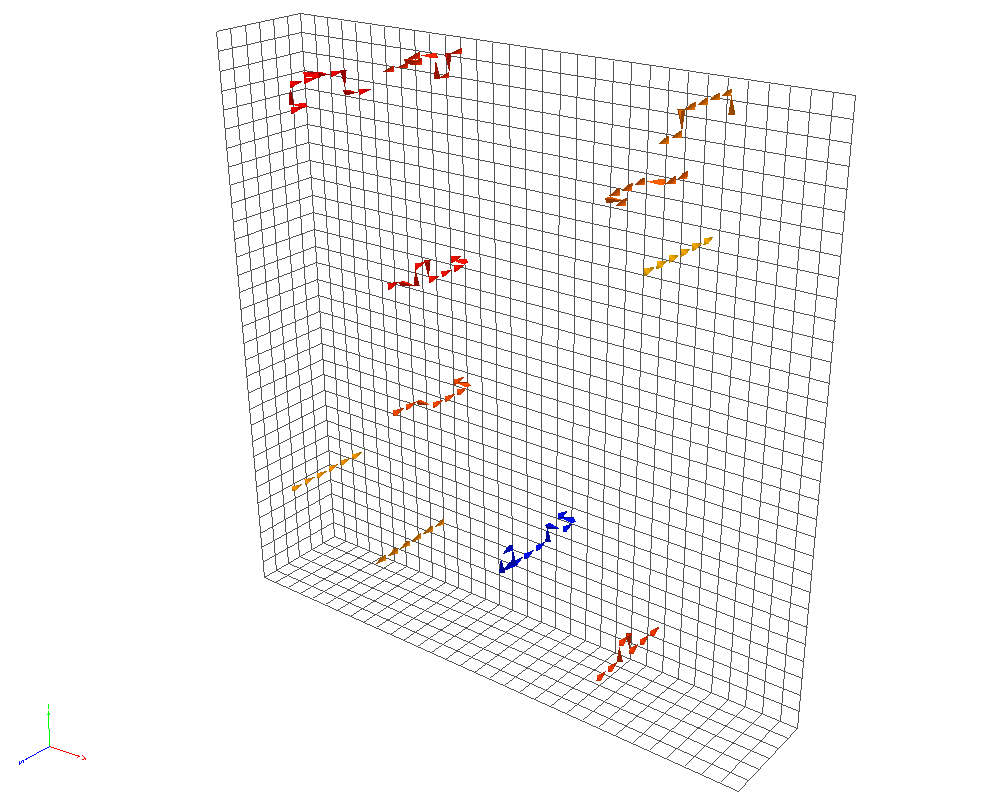}}{graphics/U3D/Nt6_spatial_slice.u3d}
\caption{Illustrations of a spatial slice of the finite temperature lattices with $N_\tau = 8$
  providing $T/T_c = 0.91$ (\textbf{upper}) and $N_\tau = 6$ providing $T/T_c = 1.22$
  (\textbf{lower}).  The blue jets illustrate the nontrivial plaquettes pierced by a vortex forming
  the percolating cluster. Secondary vortex clusters separate from the percolating cluster on this
  time slice are rendered in different colours to aid in their identification. Above $T_c$, most
  jets point in the short temporal direction, piercing spatial plaquettes and not generating
  confinement in space-time loops.}
\label{fig:spatialSlice}
\end{figure}

We now turn our attention to spatial slices of the lattice.
Recall that it is these spatial slices that expose the space-time plaquettes relevant to the Wilson
loop of Eq.~(\ref{eq:staticquarkpotential}) and confinement.  
Figure~\ref{fig:spatialSlice} provides representative visualisations of this vortex structure.

Four features are immediately apparent:
\begin{enumerate}
\addtolength{\itemsep}{-6pt}
\item A large percolating cluster is absent above $T_c$ signalling a loss of confinement.  Below
  $T_c$ a percolating vortex structure exists. Ample vortices pointing in the long spatial
  directions are available to pierce space-time Wilson loops and build an area law for
  confinement.  
\item Secondary clusters remain present below $T_c$, despite the shorter temporal dimension. The
  view ``Secondary cluster'' in the upper rendering of Fig.~\ref{fig:spatialSlice} highlights the
  10-jet vortex-chain loop that is separate from the percolating cluster in this slice.
\item Above $T_c$, the majority of vortices point in the temporal direction.  As such, they are
  associated with spatial plaquettes and do not pierce a Wilson loop to create a confining
  potential.  Thus, quark deconfinement is manifest above $T_c$.
\item The piercing of space-time plaquettes above $T_c$ is rare, indicating an alignment of the
  vortex sheet in four dimensions with the temporal axis.
\end{enumerate}

In the upper rendering of Fig.~\ref{fig:spatialSlice} there is a secondary cluster that wraps
around the periodic time extent.  The view ``Periodic secondary cluster'' highlights this
structure.  Below $T_c$, this vortex chain still pierces several space-time plaquettes, with 6 of
the 14 jets potentially contributing to a Wilson loop.  It is the straightening of this loop that
signifies the deconfinement transition..

\subsubsection{Vortex density}

Turning our attention to the quantified properties of vortex matter, an even
greater drop in the vortex density is observed in spatial slices as one crosses the critical
temperature.  Referring to Fig.~10 of Ref.~\cite{Mickley:2024zyg}, the robust first-order signal of
the phase transition is the significant $\simeq 2/3$ drop in the vortex density from $1.9$ to
$0.6\,$fm$^{-2}$.  This is manifest in Fig.~\ref{fig:spatialSlice}.

\subsubsection{Branching point density}

Figure~15 of Ref.~\cite{Mickley:2024zyg} shows a significant drop in the branching point density
across the phase transition, from values approaching $2.9\,$fm$^{-3}$ to $0.5\,$fm$^{-3}$ across
the phase transition, more than a five-fold reduction.  This is also manifest in
Fig.~\ref{fig:tempSlice}. A careful examination of the vortex chains in the lower rendering for
$T/T_c = 1.22$ shows a single pair of branching points in the blue vortex chain.  The view
``Branching point pair'' shows the two branching points with one split across the periodic
boundary.  Taking into account the $6\times 32^2$ volume and the lattice spacing of $0.10\,$fm,
this observation corresponds to a density of $0.3\,$fm$^{-3}$ suggesting that one generally finds
one to two pairs of branching points per slice.

\section{Concluding Remarks}

In this short presentation, we have revisited illustrations of vortex matter structure in pure
SU(3) gauge theory in light of the quantitative analysis of Ref.~\cite{Mickley:2024zyg} that
demonstrates the first order nature of the deconfinement phase transition in the pure-gauge theory.
Drawing on this analysis with references to the figures of Ref.~\cite{Mickley:2024zyg}, we have
highlighted specific signatures of the established properties in the representative field structure
illustrations.
In a detailed side-by-side comparison of the vortex-matter field structure slightly below and above
the critical temperature we have highlighted the vortex structure features that underpin the
quantitative results.
The abrupt step-change of the first order phase transition makes it easy to observe the field
structure changes in the visualisations and animations.

\section*{Acknowledgements}

It is a pleasure to acknowledge ongoing interesting discussions with Ryan Bignell and Chris Allton
on the role of centre vortices in QCD ground state field structure and its associated
gauge-invariant measures that quantify its properties.  We also acknowledge the contributions of
James Biddle in developing early vortex-geometry analysis techniques that continue to be of value. This
work was supported with supercomputing resources provided by the Phoenix High Performance Computing
(HPC) service at the University of Adelaide. This research was undertaken with the assistance of
resources and services from the National Computational Infrastructure (NCI), which is supported by
the Australian Government. This research was supported by the Australian Research Council through
Grant No. DP210103706. W.~K.\ was supported by the Pawsey Supercomputing Centre through the Pawsey
Centre for Extreme Scale Readiness (PaCER) program.


\begin{thebibliography}{10}

\bibitem{Mickley:2024zyg}
J.A.~Mickley, W.~Kamleh and D.B.~Leinweber, \emph{{Center vortex geometry at
  finite temperature}},
  \href{https://doi.org/10.1103/PhysRevD.110.034516}{\emph{Phys. Rev. D}
  {\bfseries 110} (2024) 034516}
  [\href{https://arxiv.org/abs/2405.10670}{{\ttfamily 2405.10670}}].

\bibitem{Mickley:2024vkm}
J.A.~Mickley, C.~Allton, R.~Bignell and D.B.~Leinweber, \emph{{Center vortex
  evidence for a second finite-temperature QCD transition}},
  \href{https://doi.org/10.1103/PhysRevD.111.034508}{\emph{Phys. Rev. D}
  {\bfseries 111} (2025) 034508}
  [\href{https://arxiv.org/abs/2411.19446}{{\ttfamily 2411.19446}}].

\bibitem{tHooft:1977nqb}
G.~'t~Hooft, \emph{{On the Phase Transition Towards Permanent Quark
  Confinement}},
  \href{https://doi.org/10.1016/0550-3213(78)90153-0}{\emph{Nucl. Phys. B}
  {\bfseries 138} (1978) 1}.

\bibitem{tHooft:1979rtg}
G.~'t~Hooft, \emph{{A Property of Electric and Magnetic Flux in Nonabelian
  Gauge Theories}},
  \href{https://doi.org/10.1016/0550-3213(79)90595-9}{\emph{Nucl. Phys. B}
  {\bfseries 153} (1979) 141}.

\bibitem{Nielsen:1979xu}
H.B.~Nielsen and P.~Olesen, \emph{{A Quantum Liquid Model for the QCD Vacuum:
  Gauge and Rotational Invariance of Domained and Quantized Homogeneous Color
  Fields}}, \href{https://doi.org/10.1016/0550-3213(79)90065-8}{\emph{Nucl.
  Phys. B} {\bfseries 160} (1979) 380}.

\bibitem{DelDebbio:1996lih}
L.~Del~Debbio, M.~Faber, J.~Greensite and S.~Olejnik, \emph{{Center dominance
  and Z(2) vortices in SU(2) lattice gauge theory}},
  \href{https://doi.org/10.1103/PhysRevD.55.2298}{\emph{Phys. Rev. D}
  {\bfseries 55} (1997) 2298}
  [\href{https://arxiv.org/abs/hep-lat/9610005}{{\ttfamily hep-lat/9610005}}].

\bibitem{Faber:1997rp}
M.~Faber, J.~Greensite and S.~Olejnik, \emph{{Casimir scaling from center
  vortices: Towards an understanding of the adjoint string tension}},
  \href{https://doi.org/10.1103/PhysRevD.57.2603}{\emph{Phys. Rev. D}
  {\bfseries 57} (1998) 2603}
  [\href{https://arxiv.org/abs/hep-lat/9710039}{{\ttfamily hep-lat/9710039}}].

\bibitem{DelDebbio:1998luz}
L.~Del~Debbio, M.~Faber, J.~Giedt, J.~Greensite and S.~Olejnik,
  \emph{{Detection of center vortices in the lattice Yang-Mills vacuum}},
  \href{https://doi.org/10.1103/PhysRevD.58.094501}{\emph{Phys. Rev. D}
  {\bfseries 58} (1998) 094501}
  [\href{https://arxiv.org/abs/hep-lat/9801027}{{\ttfamily hep-lat/9801027}}].

\bibitem{Montero:1999by}
A.~Montero, \emph{{Study of SU(3) vortex - like configurations with a new
  maximal center gauge fixing method}},
  \href{https://doi.org/10.1016/S0370-2693(99)01113-2}{\emph{Phys. Lett. B}
  {\bfseries 467} (1999) 106}
  [\href{https://arxiv.org/abs/hep-lat/9906010}{{\ttfamily hep-lat/9906010}}].

\bibitem{Bertle:1999tw}
R.~Bertle, M.~Faber, J.~Greensite and S.~Olejnik, \emph{{The Structure of
  projected center vortices in lattice gauge theory}},
  \href{https://doi.org/10.1088/1126-6708/1999/03/019}{\emph{JHEP} {\bfseries
  03} (1999) 019} [\href{https://arxiv.org/abs/hep-lat/9903023}{{\ttfamily
  hep-lat/9903023}}].

\bibitem{Faber:1999gu}
M.~Faber, J.~Greensite, S.~Olejnik and D.~Yamada, \emph{{The Vortex finding
  property of maximal center (and other) gauges}},
  \href{https://doi.org/10.1088/1126-6708/1999/12/012}{\emph{JHEP} {\bfseries
  12} (1999) 012} [\href{https://arxiv.org/abs/hep-lat/9910033}{{\ttfamily
  hep-lat/9910033}}].

\bibitem{Engelhardt:1999xw}
M.~Engelhardt and H.~Reinhardt, \emph{{Center projection vortices in continuum
  Yang-Mills theory}},
  \href{https://doi.org/10.1016/S0550-3213(99)00727-0}{\emph{Nucl. Phys. B}
  {\bfseries 567} (2000) 249}
  [\href{https://arxiv.org/abs/hep-th/9907139}{{\ttfamily hep-th/9907139}}].

\bibitem{Bertle:2000qv}
R.~Bertle, M.~Faber, J.~Greensite and S.~Olejnik, \emph{{P vortices, gauge
  copies, and lattice size}},
  \href{https://doi.org/10.1088/1126-6708/2000/10/007}{\emph{JHEP} {\bfseries
  10} (2000) 007} [\href{https://arxiv.org/abs/hep-lat/0007043}{{\ttfamily
  hep-lat/0007043}}].

\bibitem{Engelhardt:2003wm}
M.~Engelhardt, M.~Quandt and H.~Reinhardt, \emph{{Center vortex model for the
  infrared sector of SU(3) Yang-Mills theory: Confinement and deconfinement}},
  \href{https://doi.org/10.1016/j.nuclphysb.2004.02.036}{\emph{Nucl. Phys. B}
  {\bfseries 685} (2004) 227}
  [\href{https://arxiv.org/abs/hep-lat/0311029}{{\ttfamily hep-lat/0311029}}].

\bibitem{Langfeld:2001cz}
K.~Langfeld, H.~Reinhardt and J.~Gattnar, \emph{{Gluon propagators and quark
  confinement}},
  \href{https://doi.org/10.1016/S0550-3213(01)00574-0}{\emph{Nucl. Phys. B}
  {\bfseries 621} (2002) 131}
  [\href{https://arxiv.org/abs/hep-ph/0107141}{{\ttfamily hep-ph/0107141}}].

\bibitem{Greensite:2003bk}
J.~Greensite, \emph{{The Confinement problem in lattice gauge theory}},
  \href{https://doi.org/10.1016/S0146-6410(03)90012-3}{\emph{Prog. Part. Nucl.
  Phys.} {\bfseries 51} (2003) 1}
  [\href{https://arxiv.org/abs/hep-lat/0301023}{{\ttfamily hep-lat/0301023}}].

\bibitem{Langfeld:2003ev}
K.~Langfeld, \emph{{Vortex structures in pure SU(3) lattice gauge theory}},
  \href{https://doi.org/10.1103/PhysRevD.69.014503}{\emph{Phys. Rev.}
  {\bfseries D69} (2004) 014503}
  [\href{https://arxiv.org/abs/hep-lat/0307030}{{\ttfamily hep-lat/0307030}}].

\bibitem{OCais:2008kqh}
A.~O'Cais, W.~Kamleh, K.~Langfeld, B.~Lasscock, D.~Leinweber, P.~Moran et~al.,
  \emph{{Preconditioning Maximal Center Gauge with Stout Link Smearing in
  SU(3)}}, \href{https://doi.org/10.1103/PhysRevD.82.114512}{\emph{Phys. Rev.
  D} {\bfseries 82} (2010) 114512}
  [\href{https://arxiv.org/abs/0807.0264}{{\ttfamily 0807.0264}}].

\bibitem{Bowman:2010zr}
P.O.~Bowman, K.~Langfeld, D.B.~Leinweber, A.~Sternbeck, L.~von Smekal and
  A.G.~Williams, \emph{{Role of center vortices in chiral symmetry breaking in
  SU(3) gauge theory}},
  \href{https://doi.org/10.1103/PhysRevD.84.034501}{\emph{Phys. Rev. D}
  {\bfseries 84} (2011) 034501}
  [\href{https://arxiv.org/abs/1010.4624}{{\ttfamily 1010.4624}}].

\bibitem{OMalley:2011aa}
E.-A.~O'Malley, W.~Kamleh, D.~Leinweber and P.~Moran, \emph{{SU(3) centre
  vortices underpin confinement and dynamical chiral symmetry breaking}},
  \href{https://doi.org/10.1103/PhysRevD.86.054503}{\emph{Phys. Rev. D}
  {\bfseries 86} (2012) 054503}
  [\href{https://arxiv.org/abs/1112.2490}{{\ttfamily 1112.2490}}].

\bibitem{Trewartha:2015nna}
A.~Trewartha, W.~Kamleh and D.~Leinweber, \emph{{Evidence that centre vortices
  underpin dynamical chiral symmetry breaking in SU(3) gauge theory}},
  \href{https://doi.org/10.1016/j.physletb.2015.06.025}{\emph{Phys. Lett.}
  {\bfseries B747} (2015) 373}
  [\href{https://arxiv.org/abs/1502.06753}{{\ttfamily 1502.06753}}].

\bibitem{Trewartha:2015ida}
A.~Trewartha, W.~Kamleh and D.~Leinweber, \emph{{Connection between center
  vortices and instantons through gauge-field smoothing}},
  \href{https://doi.org/10.1103/PhysRevD.92.074507}{\emph{Phys. Rev.}
  {\bfseries D92} (2015) 074507}
  [\href{https://arxiv.org/abs/1509.05518}{{\ttfamily 1509.05518}}].

\bibitem{Greensite:2016pfc}
J.~Greensite, \emph{{Confinement from Center Vortices: A review of old and new
  results}}, \href{https://doi.org/10.1051/epjconf/201713701009}{\emph{EPJ Web
  Conf.} {\bfseries 137} (2017) 01009}
  [\href{https://arxiv.org/abs/1610.06221}{{\ttfamily 1610.06221}}].

\bibitem{Trewartha:2017ive}
A.~Trewartha, W.~Kamleh and D.~Leinweber, \emph{{Centre vortex removal restores
  chiral symmetry}}, \href{https://doi.org/10.1088/1361-6471/aa9443}{\emph{J.
  Phys.} {\bfseries G44} (2017) 125002}
  [\href{https://arxiv.org/abs/1708.06789}{{\ttfamily 1708.06789}}].

\bibitem{Biddle:2018dtc}
J.C.~Biddle, W.~Kamleh and D.B.~Leinweber, \emph{{Gluon propagator on a
  center-vortex background}},
  \href{https://doi.org/10.1103/PhysRevD.98.094504}{\emph{Phys. Rev.}
  {\bfseries D98} (2018) 094504}
  [\href{https://arxiv.org/abs/1806.04305}{{\ttfamily 1806.04305}}].

\bibitem{Biddle:2019gke}
J.C.~Biddle, W.~Kamleh and D.B.~Leinweber, \emph{{Visualization of center
  vortex structure}},
  \href{https://doi.org/10.1103/PhysRevD.102.034504}{\emph{Phys. Rev. D}
  {\bfseries 102} (2020) 034504}
  [\href{https://arxiv.org/abs/1912.09531}{{\ttfamily 1912.09531}}].

\bibitem{Biddle:2020isk}
J.C.~Biddle, W.~Kamleh and D.B.~Leinweber, \emph{{Emergent Structure in QCD}},
  \href{https://doi.org/10.1051/epjconf/202024506009}{\emph{EPJ Web Conf.}
  {\bfseries 245} (2020) 06009}
  [\href{https://arxiv.org/abs/2009.12044}{{\ttfamily 2009.12044}}].

\bibitem{Biddle:2022acd}
J.C.~Biddle, W.~Kamleh and D.B.~Leinweber, \emph{{Impact of dynamical fermions
  on the center vortex gluon propagator}},
  \href{https://doi.org/10.1103/PhysRevD.106.014506}{\emph{Phys. Rev. D}
  {\bfseries 106} (2022) 014506}
  [\href{https://arxiv.org/abs/2206.02320}{{\ttfamily 2206.02320}}].

\bibitem{Biddle:2022zgw}
J.C.~Biddle, W.~Kamleh and D.B.~Leinweber, \emph{{Static quark potential from
  center vortices in the presence of dynamical fermions}},
  \href{https://doi.org/10.1103/PhysRevD.106.054505}{\emph{Phys. Rev. D}
  {\bfseries 106} (2022) 054505}
  [\href{https://arxiv.org/abs/2206.00844}{{\ttfamily 2206.00844}}].

\bibitem{Virgili:2022ybm}
A.~Virgili, W.~Kamleh and D.~Leinweber, \emph{{Smoothing algorithms for
  projected center-vortex gauge fields}},
  \href{https://doi.org/10.1103/PhysRevD.106.014505}{\emph{Phys. Rev. D}
  {\bfseries 106} (2022) 014505}
  [\href{https://arxiv.org/abs/2203.09764}{{\ttfamily 2203.09764}}].

\bibitem{Biddle:2023lod}
J.C.~Biddle, W.~Kamleh and D.B.~Leinweber, \emph{{Center vortex structure in
  the presence of dynamical fermions}},
  \href{https://doi.org/10.1103/PhysRevD.107.094507}{\emph{Phys. Rev. D}
  {\bfseries 107} (2023) 094507}
  [\href{https://arxiv.org/abs/2302.05897}{{\ttfamily 2302.05897}}].

\bibitem{Kamleh:2023gho}
W.~Kamleh, D.B.~Leinweber and A.~Virgili, \emph{{Numerical indication that
  center vortices drive dynamical mass generation in QCD}},
  \href{https://doi.org/10.1103/PhysRevD.110.L051502}{\emph{Phys. Rev. D}
  {\bfseries 110} (2024) L051502}
  [\href{https://arxiv.org/abs/2305.18690}{{\ttfamily 2305.18690}}].

\bibitem{Mickley:2025ksp}
J.A.~Mickley, D.B.~Leinweber and L.E.~Oxman, \emph{{Structure of center vortex
  matter in SU(4) Yang-Mills theory}},
  \href{https://arxiv.org/abs/2502.19656}{{\ttfamily 2502.19656}}.

\bibitem{Leinweber:2022ukj}
D.~Leinweber, J.~Biddle, W.~Kamleh and A.~Virgili, \emph{{Dynamical fermions,
  centre vortices, and emergent phenomena}},
  \href{https://doi.org/10.1051/epjconf/202227401002}{\emph{EPJ Web Conf.}
  {\bfseries 274} (2022) 01002}
  [\href{https://arxiv.org/abs/2211.13421}{{\ttfamily 2211.13421}}].

\bibitem{Gross:2022hyw}
F.~Gross et~al., \emph{{50 Years of Quantum Chromodynamics}},
  \href{https://doi.org/10.1140/epjc/s10052-023-11949-2}{\emph{Eur. Phys. J. C}
  {\bfseries 83} (2023) 1125}
  [\href{https://arxiv.org/abs/2212.11107}{{\ttfamily 2212.11107}}].

\bibitem{Iwasaki:1996sn}
Y.~Iwasaki, K.~Kanaya, T.~Kaneko and T.~Yoshie, \emph{{Scaling in SU(3) pure
  gauge theory with a renormalization group improved action}},
  \href{https://doi.org/10.1103/PhysRevD.56.151}{\emph{Phys. Rev. D} {\bfseries
  56} (1997) 151} [\href{https://arxiv.org/abs/hep-lat/9610023}{{\ttfamily
  hep-lat/9610023}}].

\end{thebibliography}

\providecommand{\href}[2]{#2}\begingroup\raggedright\endgroup

\end{document}